\documentclass[mathleft]{an}

\usepackage{graphicx}
\usepackage{amsfonts,amssymb}
\usepackage{times}

\newcommand{\Rey}{\mathrm{Re}}

\newcommand{\Rin}{R_\mathrm{in}}
\newcommand{\Rout}{R_\mathrm{out}}

\newcommand{\Pm}{\mathrm{Pm}}
\newcommand{\Ha}{\mathrm{Ha}}

\newcommand{\mperm}{\mu_0}
\newcommand{\mdiff}{\eta}
\newcommand{\cmnt}[1]{}
\newcommand{\comm}[1]{}
\newcommand{\ignore}[1]{}
\def\rot{\mathop{\rm rot}\nolimits}
\def\div{\mathop{\rm div}\nolimits} 
\def\gsim{\lower.4ex\hbox{$\;\buildrel >\over{\scriptstyle\sim}\;$}} 
\def\lsim{\lower.4ex\hbox{$\;\buildrel <\over{\scriptstyle\sim}\;$}} 

\def\qq{\qquad\qquad}                      
\def\q{\qquad}
\def\beg{\begin{eqnarray}}
\def\ende{\end{eqnarray}}
\renewcommand{\vec}[1]{\mbox{\boldmath $#1$}}

\newcommand{\Om}{{\it \Omega}}

\begin{document}

\sloppy

 \Pagespan{1}{7}
 \Yearpublication{2011}%
 \Yearsubmission{2010}%
 \Month{11}%
 \Volume{XXX}%
 \Issue{X}%
 \DOI{}

 \title{The Tayler instability of toroidal magnetic fields in a columnar   gallium  experiment}

 \author{G. R\"udiger$^{1,2,}$\thanks{Corresponding author:
   gruediger@aip.de} \and M. Schultz$^1$ \and M. Gellert$^1$}

 \institute{
 Astrophysikalisches Institut Potsdam, An der Sternwarte 16, 
 D-14482 Potsdam, Germany \and  Forschungszentrum Dresden Rossendorf, P.O. Box 510119, D-01314 Dresden, Germany} 
 \authorrunning{G. R\"udiger, M. Schultz\and M. Gellert}
 \titlerunning{Tayler instability of toroidal magnetic fields} 
 \received{2010}
 \accepted{2010}
 \publonline{XXX}


 \abstract{The nonaxisymmetric  Tayler instability  of toroidal magnetic fields due to axial electric currents is studied for conducting incompressible fluids between two coaxial cylinders  without endplates. The inner cylinder is considered as so  thin that even the limit of $R_{\rm in} \to 0$  can be computed.  The magnetic Prandtl number is varied over many orders of magnitudes but the azimuthal mode number of the perturbations is fixed to $m=1$. In the linear approximation the critical magnetic field amplitudes and the growth rates of the instability are determined for both resting and  rotating cylinders. Without rotation the critical Hartmann numbers do {\em not} depend on the magnetic Prandtl number but this is not true for the growth rates. For given product of viscosity and magnetic diffusivity the growth rates for small and large magnetic Prandtl number are much smaller than those for $\rm Pm=1$. For gallium under the influence of a  magnetic field at the outer cylinder of 1\,kG the resulting growth time is 5\,s. The minimum electric current through  a container of 10 cm diameter to excite the kink-type instability is 3.20 kA. For a rotating container  both the critical magnetic field and the related growth times are larger than for the resting column.}

 \maketitle

 \section{Motivation}
 A known  instability of toroidal fields is the current-driven (`kink-type') Tayler  instability (TI)  
 which is basically nonaxisymmetric (Tayler 1957; Vandakurov 1972; Tayler 1973; Acheson 1978). The toroidal 
 field becomes unstable against nonaxisymmetric perturbations for a sufficiently large magnetic field 
 amplitude depending on the radial profile of the field.  A global rigid rotation of the system stabilizes the TI, i.e. 
 much higher   field amplitudes can be kept stable. For the rapidly rotating 
 regime $\Om>\Om_{\rm A}$ (with $\Om_{\rm A}$ is the Alfv\'en frequency of the toroidal field) 
 the stability becomes complete, i.e. {\em all} possible modes in incompressible fluids of 
 uniform density are stable (Pitts \& Tayler 1985). 
 We shall demonstrate in the present paper how this instability and its stabilization by 
 rigid rotation  can experimentally be realized with fluid conductors like sodium or gallium. 
 There is so far no empirical or observational proof of the 
existence of the TI (Maeder \& Meynet 2005).

The stability problem of a system of toroidal fields and differential rotation has been studied in cylindric geometry (Michael 1954; Chandrasekhar 1961; Howard \& Gupta 1962; Chanmugam 1979; Knobloch 1992: Dubrulle \& Knobloch 1993; Kumar, Coleman \& Kley 1994;  Pessah \& Psaltis 2005; Shalybkov 2006) but in   all   studies only  axi\-symmetric perturbations are considered.     Here  as a continuation of the papers by R\"udiger et al. (2007a,b) and  R\"udiger \& Schultz (2010)
the  attention is focused to the nonaxisymmetric perturbation  modes with ${m=1}$
for real fluids. In particular, the possible  realizations of the instabilities
as an experiment laboratory in progress  are discussed.

\section{The Taylor-Couette geometry}
A Taylor-Couette container is considered  confining  a  toroidal magnetic field of given amplitudes at the cylinders which rotate with the common rotation rate  $\Om$. We are here only  interested in the hydrodynamically
 stable regime with rigid rotation. The gap between the cylinders is considered as so wide that also  a column without any inner cylinder can be simulated.  Formally, the inner radius 
is  $100 \hat\eta$\,\% of the outer radius. 
Extreme values of $\hat\eta=0.001$  are used in the present paper contrary to the calculations by R\"udiger et al. (2007a) for a medium gap of $\hat\eta=0.5$.
The fluid  between the cylinders is assumed to be incompressible with uniform density and
 dissipative with the kinematic viscosity $\nu$ and the magnetic diffusivity
 $\mdiff$. 

The container possesses the inner and outer cylinder with radii $R_{\rm in}$ and $R_{\rm out}$ and with their ratio
\beg
\hat\eta=\frac{R_{\rm{in}}}{R_{\rm{out}}}.
\label{mu}
\ende

The radial profiles of the magnetic background fields are restricted for real fluids. The solution of the stationary induction equation without  shear reads 
\beg
B_\phi=A R+\frac{B}{R}
\label{basic}
\ende
in the cylinder geometry.
 $A$ and $B$  are the basic parameters;  the term $A R$  corresponds to  uniform axial
currents  with $I=2A$ everywhere within $R<R_{\rm out}$, and $B/R$ corresponds to
a uniform additional electric current only within $R<R_{\rm in}$. In the present paper we generally put 
\beg
B=0
\label{B}
\ende
 (see Roberts 1956; Pitts \& Tayler 1985) with 
the consequence that the azimuthal magnetorotational instability (`AMRI', see R\"udiger 2007a) does not appear.   The  behavior 
of the toroidal field is thus only due to the TI for magnetic fields  increasing outwards. 
It is useful to define the quantity
\beg
\mu_B=\frac{B_{\rm{out}}}{B_{\rm{in}}},
\ende
measuring the variation of $B_\phi$ across the gap. Vanishing $B$ leads to   $\mu_B=1/\hat\eta$. 
In the following we  fix  $\mu_B=1/\hat\eta$ but we shall vary the magnetic Prandtl number
\beg
\rm Pm=\frac{\nu}{\eta},
\label{pm}
\ende
and also the (small) values of $\hat\eta$. 
\section{Equations and numerical model}
The dimensionless  MHD equations for incompressible fluids are
 \begin{eqnarray}
 \lefteqn{   {\rm Re}    \frac{\partial\vec{u}}{\partial t} +{\rm Re} (\vec{u} \cdot \nabla)\vec{u} =
               -\nabla P + \Delta \vec{u} + {\Ha^2}
               \rot \vec{B} \times \vec{B},} \nonumber\\
\lefteqn{   {\rm Rm}    \frac{\partial \vec{B}}{\partial t} =
                 \Delta \vec{B} + {\rm Rm}
	        \rot (\vec{u} \times \vec{B}),} 
\label{7}
\end{eqnarray}
with       $\div{\vec{u}} =  \div{\vec{B}} = 0$ and
  with the Hartmann number
 \beg
    \Ha = \frac{B_{\rm in} D}{\sqrt{\mperm \rho \nu \mdiff}}.
 \label{Hart}
 \ende
 Here  $D=\sqrt{\Rin(\Rout-\Rin)}$ is used as the unit of length,  $\eta/D$ as the 
 unit of velocity and $B_{\rm in}$ as the unit of magnetic fields. Frequencies including the  rotation rate $\Om$ are normalized with the inner rotation rate $\Om_{\rm in}$. The Reynolds number
 $\Rey$ is defined as 
 \beg
 \Rey=\frac{\Om  D^2}{ \nu},
 \label{re}
 \ende
  and the magnetic Reynolds number is 
\beg
{\rm Rm}=\frac{\Om D^2}{\eta}.
\label{10}
\ende 
Sometimes it appears   as useful to work with the `mixed' Reynolds number 
 \beg
{\rm Rm^*=\sqrt{\rm Re \cdot Rm}} 
\label{rem1}
\ende
which is symmetric in $\nu$ and $\eta$ as it is also the Hartmann number.  
For ${\rm Pm}=1$ it is ${\rm Re}={\rm Rm}={\rm Rm^*}$. It is also useful to use the Lundquist number ${\rm S}= \sqrt{\rm Pm}\ {\rm Ha}$. 
The ratio of $\rm Rm^*$ and Ha, 
\beg
{\rm Mm}= \frac{{\rm Rm^*}}{{\rm Ha}},
\label{Mach}
\ende
is called the magnetic Mach number of rotation.

Applying the usual normal mode analysis, we look for solutions of the
linearized equations of the form
\beg
F=F(R)\,{\rm{exp}}\bigl[{\rm{i}}(kz+m\phi+\omega t)\bigr].
\label{nmode}
\ende
Using Eq. (\ref{nmode}), linearizing the Eq. (\ref{7})
and representing the result as a system of first order equations, one finds
\begin{eqnarray}
\lefteqn{\frac{{\rm d}u_R}{{\rm d}R}+\frac{u_R}{R}+{\textrm i}\frac{m}{R}u_\phi+{\textrm i}ku_z=0,}
\nonumber \\
\lefteqn{\frac{{\rm d}P}{{\rm d}R}+{\textrm i}\frac{m}{R}\phi_u+{\textrm i}kZ
+\left(k^2+\frac{m^2}{R^2}\right)u_R\ +}
\nonumber \\
&& \qq \q +\ {\textrm{iRe}}(\omega+m\Om)u_R-2\Om {\textrm{Re}} u_\phi- 
\nonumber \\
&& \qq \q -\ {\textrm{iHa}}^2 mA b_R +2{\textrm{Ha}}^2 A
b_\phi=0,
\nonumber \\
\lefteqn{\frac{{\rm d}\phi_u}{{\rm d}R}-\left(k^2+\frac{m^2}{R^2}\right)u_\phi-
{\textrm{iRe}}(\omega+m\Om)u_\phi\ +}
\nonumber \\
&& \qq \q +\ 2{\textrm i}\frac{m}{R^2}u_R
- 2{\textrm{Re}}\Om u_R+
\nonumber \\
&&\qq \q +\ 2{\textrm{Ha}}^2A b_R
+{\textrm{iHa}}^2 mA b_\phi
-{\textrm{i}}\frac{m}{R}P=0,
\nonumber \\
\lefteqn{\frac{{\rm d}Z}{{\rm d}R}+\frac{Z}{R}-\left(k^2+\frac{m^2}{R^2}\right)u_z-
{\textrm{iRe}}(\omega+m\Om)u_z\ -}
\nonumber \\
&& \qq \q -\ {\textrm i}kP+{\textrm{iHa}}^2mA b_z=0,
\nonumber \\
\lefteqn{\frac{{\rm d}b_R}{{\rm d}R}+\frac{b_R}{R}+{\textrm i}\frac{m}{R}b_\phi+{\textrm i}kb_z=0,}
\nonumber \\
\lefteqn{\frac{{\rm d}b_z}{{\rm d}R}-\frac{{\textrm i}}{k}\left(k^2+\frac{m^2}{R^2}\right)b_R
+{\textrm{PmRe}}\frac{1}{k}(\omega+m\Om)b_R\ +}
\nonumber \\
&& \qq \q +\ \frac{1}{k}\frac{m}{R}\phi_B-\frac{1}{k}mA u_R=0,
\nonumber \\
\lefteqn{\frac{{\rm d}\phi_B}{{\rm d}R}-\left(k^2+\frac{m^2}{R^2}\right)b_\phi
-{\textrm{iPmRe}}(\omega+m\Om)b_\phi\ +}
\nonumber \\
&& \qq \q +\ {\textrm i}\frac{2m}{R^2}b_R
-R u_R
 +{\textrm i}mA u_\phi =0,
\label{syst}
\end{eqnarray}
where $\phi_u$, $Z$ and $\phi_B$ are defined as
\begin{equation}
\phi_u=\frac{{\rm d}u_\phi}{{\rm d}R}+\frac{u_\phi}{R},
\ \ \ \
Z=\frac{{\rm d}u_z}{{\rm d}R}, \ \ \ \ 
\phi_B=\frac{{\rm d}b_\phi}{{\rm d}R}+\frac{b_\phi}{R},
\label{def}
\end{equation}
and $A=1/R_{\rm in}$ ($R_{\rm in}$ in units of $D$).

 An appropriate set of ten boundary conditions is needed to solve the
system (\ref{syst}).  For the velocity the boundary conditions are always
no-slip while the cylinders are assumed to be perfect conductors.
These boundary conditions are applied at both $R_{\rm{in}}$ and $R_{\rm{out}}$. The wave number 
is varied as long as  for given Hartmann number the Reynolds number takes its minimum. 
The procedure is  described in 
detail by Shalybkov, R\"udiger \& Schultz (2002). One immediately finds that the sign of the real wave 
number $k$ is free so that with the solution for $k$ also another one  with $-k$ exists.
For insulating walls the boundary conditions are  more complicated (see R\"udiger et al. 2007b).

 Tayler (1973) found the necessary and sufficient condition
\beg
\frac{{\rm d}}{{\rm d}R}\left(RB_\phi^2\right) < 0
\label{ddR}
\ende
for   stability of an ideal nonrotating fluid against nonaxisymmetric disturbances. Our field profile $B_\phi=AR$  is thus stable against $m=0$ and unstable for sufficiently large field amplitudes, i.e. for ${\rm Ha}>{\rm Ha}_{\rm crit}$. The same is true for the  nearly uniform fields with $\mu_B \simeq 1$. A general  criterion for rotating flows does not exist.

The system (\ref{syst}) has been used  to demonstrate that for resting  cylinders the ${\rm Ha}_{\rm crit}$ do not depend on the magnetic Prandtl number $\rm Pm$ (R\"udiger \& Schultz 2010). 
Shalybkov (2006) has given a similar result  for $m=0$.

 \section{Resting container}
In the present paper the radius of the inner  cylinders is assumed as very small or even vanishing ($\hat\eta\ll 1$).   We have recently  shown with similar models that indeed  wide gaps are much more suitable for TI  experiments with liquid metals like sodium or gallium than narrow gaps. The nonaxisymmetric modes which are excited by the Tayler instability in resting columnar  containers do {\em not} rotate as their 
drift frequencies $\Re{(\omega)}$ vanish. 
\subsection{Critical magnetic field amplitudes}
\begin{figure}
\vskip-3mm
\vbox{  
 \includegraphics[width=8.5cm,height=5.0cm]{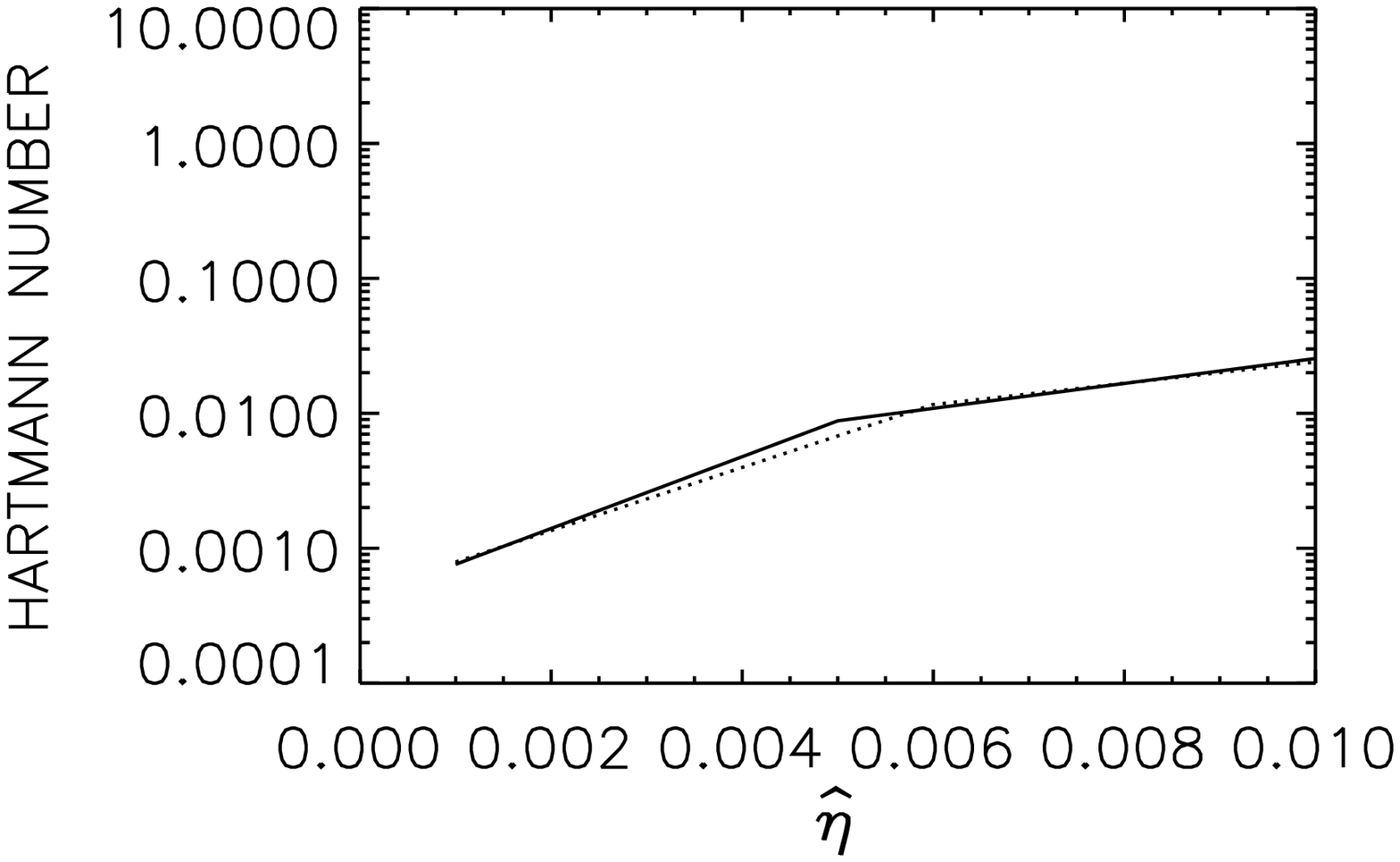} \vskip-4mm
 \includegraphics[width=8.5cm,height=5.0cm]{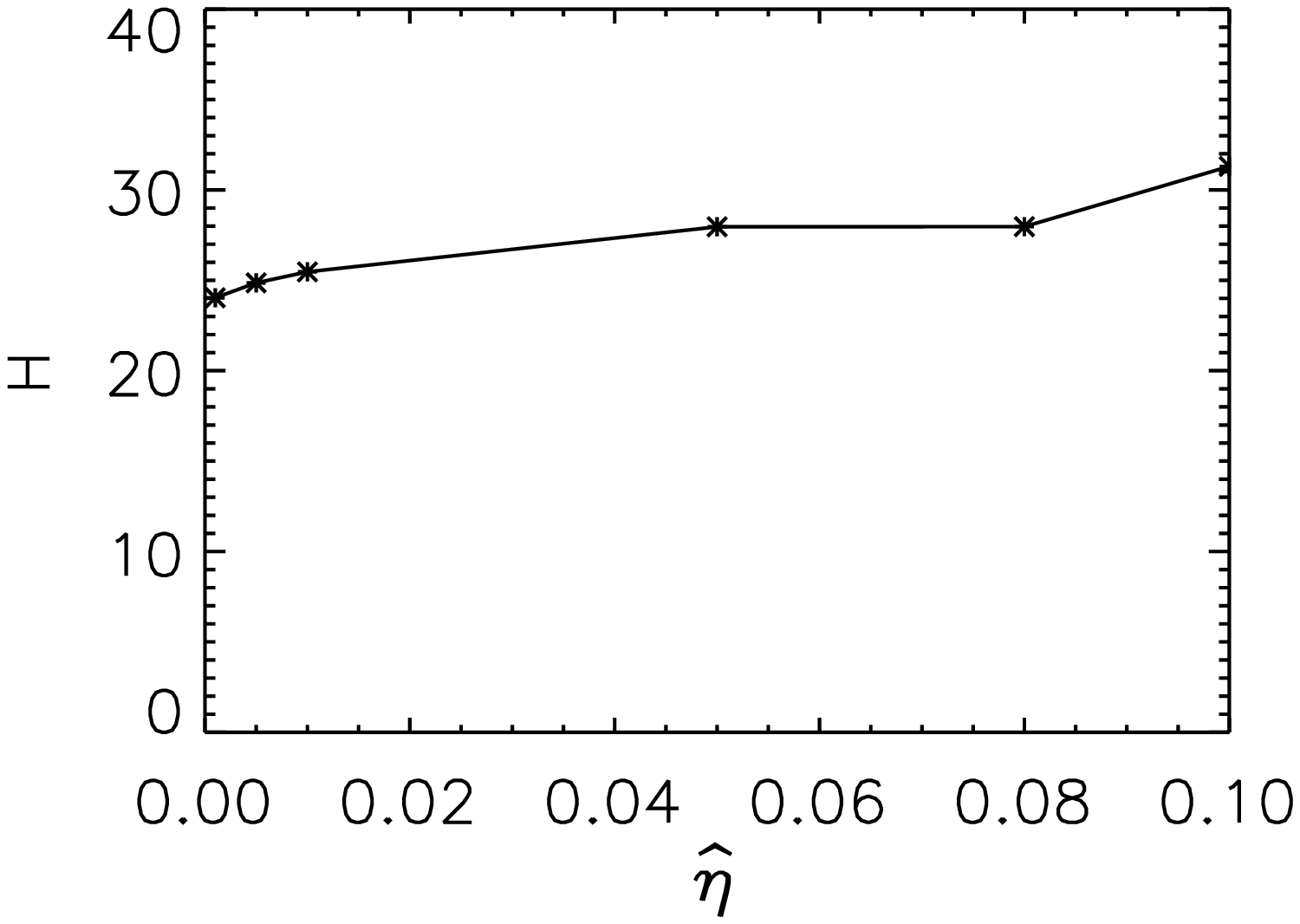}}
\vskip-2mm
\caption{\emph{Top}: the critical Hartmann number   for very  wide container gaps 
(${\hat\eta\to 0}$) between conducting and resting cylinders.  The dotted line
represents the profile (\ref{Heta}) with  ${H=25}$.  \emph{Bottom}: note the high quality of the representation (\ref{Heta}). Valid for all magnetic Prandtl numbers.}  
    \label{Ha1}
 \end{figure}

The resulting Hartmann numbers ${\rm Ha}_{\rm crit}$ for various $\hat\eta$ are given in Fig. \ref{Ha1}. They vary over many orders of magnitude and become very small for wide gaps ($\hat\eta\to 0$). For small $\hat\eta$  the critical Hartmann number vanishes like
\beg
{\rm Ha}_{\rm crit} = H\, {\hat\eta}^{1.5}.
\label{Heta}
\ende
Our calculations with the smallest  $R_{\rm in}$ concern  $\hat\eta=0.001$  and provide  $\rm Ha_{\rm crit}= 0.00075$ which leads to   $H\simeq  25$ (see Fig. \ref{Ha1}, bottom). For given $H$   the relation (\ref{Heta}) simply yields
\beg
B_{\rm out} = \frac{H}{R_{\rm out}}  \sqrt{\rho\mu_0 \nu\eta}.
\label{H}
\ende
for the critical magnetic field at the outer cylinder. 
Hence, the technical constraints  to realize the TI  in the laboratory working with liquid metals are the following.
For a nonrotating container with ${R_{\rm in}\to 0}$ and ${R_{\rm out}=5}$\ cm and  for  the gallium-tin alloy used in the experiment PROMISE ($\sqrt{\rho\mu_0 \nu\eta}=25.7$ in c.g.s.), Eq. (\ref{H}) leads to ${B_{\rm out}\simeq 128}$\ G. Now let $I$ be the axial current inside  the container, i.e. 
\beg
B_{\rm out}=\frac{I}{5R_{\rm out}},
\label{Bin}
\ende
where $R$, $B$ and $I$ are measured  in cm, G, and A.
 Hence, for $R_{\rm in}\to 0$ the total current through the fluid conductor becomes 3.20 kA which is surprisingly small. We have thus shown that it should  be possible to realize the nonaxisymmetric current-driven TI in the MHD laboratory also with conducting  fluids of small magnetic Prandtl number such as  sodium and  gallium.
\subsection{Growth rates}
\begin{figure}
\vskip-5mm
 \hskip-1cm
 \vbox
{
\includegraphics[width=9.5cm,height=10.0cm]{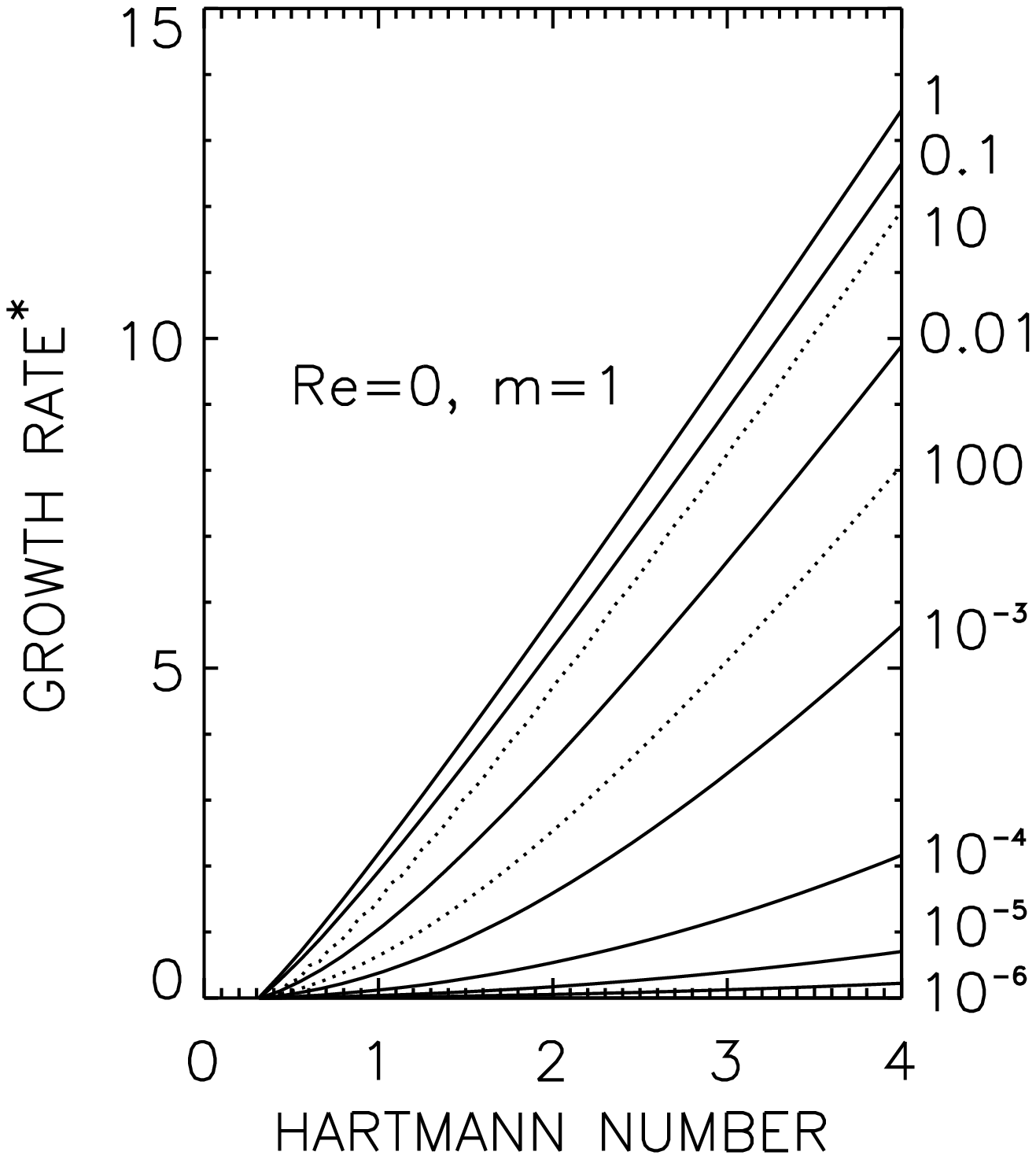} \vskip-4mm
\includegraphics[width=8.5cm,height=5.0cm]{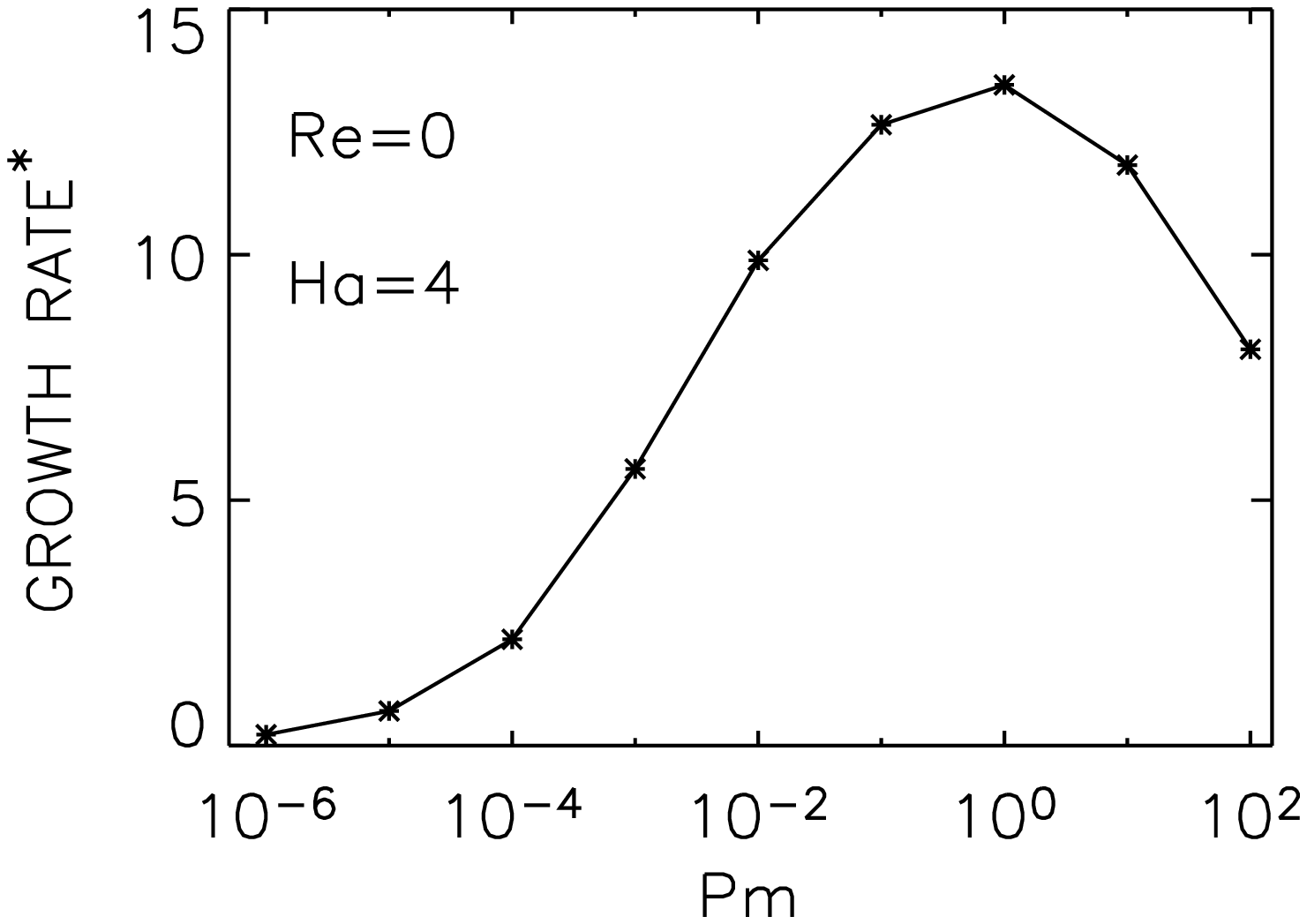}}
\vskip-2mm
\caption{\emph{Top}: the growth rates of the instability for the resting container 
normalized with the diffusion time formed with the averaged diffusivity
${\eta^*=\sqrt{\nu\eta}}$.  Fluids (or simulations) with fixed $\eta^*$ have
maximum growth rates if $\rm Pm=1$. On the right vertical axis the curves are marked with their value of $\rm Pm$. \emph{Bottom}: the same but for the fixed Hartmann number $\rm Ha=4$. The $\omega^*$ for $\rm Pm=10^{-6}$ is 0.2. It is  $\hat\eta=0.05$, $\mu_B=20$.}  
    \label{pm1}
 \end{figure}

The other important parameter for the experiment is the growth rate of the instability. Contrary to the critical Hartmann numbers the growth rates for supercritical Hartmann numbers {\em do } depend on the magnetic Prandtl number. As it is not a mild dependence over many orders of magnitude of $\rm Pm$ a rather compact formulation is needed. We prefer a normalization of frequencies 
with the averaged  diffusivity $\eta^*=\sqrt{\nu\eta}$ similar to the normalization of the magnetic field by the Hartmann number. Figure \ref{pm1} shows the normalized growth rates 
\beg
\omega^*=\frac{\omega D^2}{\eta^*}
\label{ost}
\ende
for various $\rm Ha$ and $\rm Pm$. Note that even in this formulation the growth rates strongly vary over the range $\rm Pm= 1\ldots 10^{-6}$. Fluids with equal values of $\nu$ and $\eta$ are most unstable for fixed value of $\eta^*$  ($\eta^*=2.9$\ cm$^2$/s for  Gallistan, see below). The lower plot of Fig. 
\ref{pm1} gives the values for various $\rm Pm$ but for fixed $\rm Ha$. The curve takes its maximum for $\rm Pm=1$. Note that the growth rates for very small magnetic Prandtl numbers such as $\rm Pm=10^{-6}$ are two orders of magnitudes smaller than they are for $\rm Pm=1$. 

On the first view one takes from Fig. \ref{pm1} that $\omega^*\simeq \gamma {\rm Ha}$ which yields the relation $\omega_{\rm gr}\simeq \gamma \sqrt{\hat\eta} \Om_{\rm A}$ for the physical growth rate. Here the Alfv\'{e}n frequency $\Om_{\rm A}$ is defined by 
\begin{equation}
\Om_{\rm A}=\frac{ B_{\rm out}}{\sqrt{\mu_0\rho} R_{\rm out}},
\label{20}
\end{equation}
 and it is $\gamma \lsim 4$ for ${\rm Pm}=1$. Hence, for $B_{\rm out}=1$ kGauss
 and an outer radius of (say)  $R_{\rm out} = 5$ cm the growth rate is 20
 s$^{-1}$ so that the growth time is 0.05 s. For $\rm Pm=10^{-6}$  (gallium) the
 value for $\gamma$ is only 0.05 hence the growth time for $B_{\rm out}=1$
 kG (!)  with 4 s  surprisingly long. Here and in the following the material quantities $\rho=6.4$ g/cm$^3$ and $\eta=2428$ cm$^2$/s for the gallium alloy GaInSn (``Gallistan'')  are used.

However, the profiles in Fig. \ref{pm1} are not linear. One finds that
\beg
\omega^*= \gamma_\omega\ (\rm Ha - \rm Ha_{crit})^2,
\label{ost2}
\ende
gives a better representation where $\gamma_\omega$ depends on $\rm Pm$  rather than   the value of  $\rm Ha_{crit}$. Neglecting the small values of the latter, 
\beg
\omega_{\rm gr} = \Gamma_\omega \frac{B^2_{\rm out}}{\mu_0\rho\eta} 
\label{ost2}
\ende
is obtained for the physical growth rate. Here,
\beg
\Gamma_\omega= \frac{{\hat\eta}^2 \gamma_\omega}{\sqrt{\rm Pm}}
\label{ost3}
\ende
does no longer depend on the special (small)  value of the geometry factor $\hat\eta$. Even the limit $\hat\eta\to 0$ is included.

\begin{figure}
\vskip-5mm
\hskip-3mm
  \includegraphics[width=8.5cm,height=8.0cm]{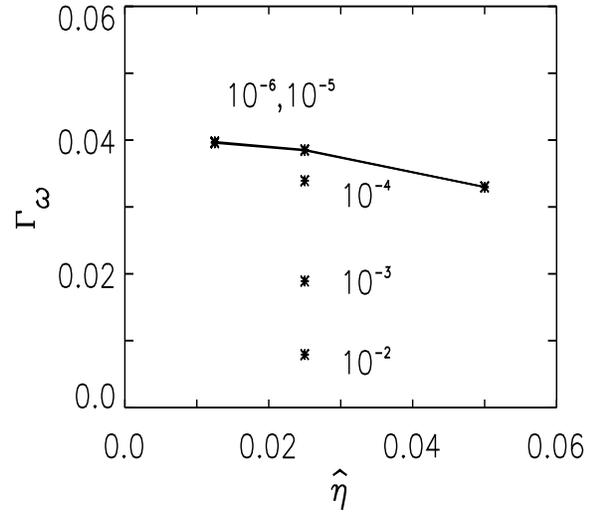}
\vskip-3mm
\caption{The parameter $\Gamma_\omega$ in Eq. (\ref{ost2}) in its dependence on the inner radius $\hat\eta$  and the magnetic Prandtl number $\rm Pm$ (marking numbers). The  $\rm Pm$-dependence disappears for small $\rm Pm$. The values hold for all 
containers with small inner radius. For $\rm Pm=1$ it is $\Gamma_\omega=0.0009 $. }
    \label{pm3}
 \end{figure}

A weak dependence on the magnetic Prandtl number remains. Figure \ref{pm3} gives the numerical values of this quantity for  various $\rm Pm$. We find the limit $\rm Pm\to 0$ as possible.
For fixed $\eta$ we now find fluids with $\rm Pm=1$ as most stable while small or large viscosities are destabilizing.

For $\rm Pm=10^{-6}$ we take from Fig.  \ref{pm3} the value $\Gamma_\omega= 0.041$. Hence,
\beg
\omega_{\rm gr}= 2.0\times\! 10^{-3} B^2_{100}\, ,
\label{om}
\ende
which yields 
\beg
\tau_{\rm gr}= \frac{500 }{B_{100}^2}
\label{time}
\ende
(with $B_{100}=B_{\rm out}/100\, {\rm G}$) as the growth time in seconds
for the TI in  gallium experiments. The size of the container does no longer
influence the growth rate. The growth time for $B_{\rm out}=1$ kG now
results to 5\,s -- and to ${\sim\! 2}$ min for, say, 200 G.


\section{Rotating container}
\begin{figure}
\vskip-6mm
\hskip-5mm
   \includegraphics[width=8.9cm,height=8.0cm]{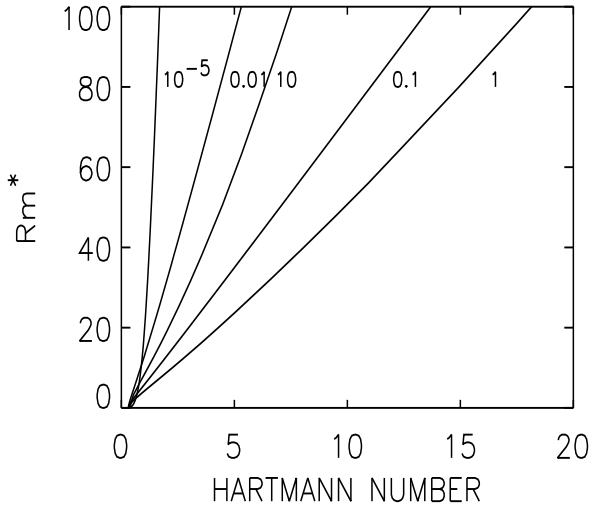}
\vskip-3mm
  \caption{The suppression of TI by uniform rotation  for $\hat\eta=0.05$. The curves are marked with their  magnetic Prandtl numbers. In this representation  fluids  with $\rm Pm=1$ undergo the  strongest stabilization by rotation. The critical Hartmann number for resting container is 0.31 for all Pm.}  
    \label{rem}
 \end{figure}

Figure \ref{rem} gives for the wide-gap container the rotational quenching of the TI for various values of the magnetic Prandtl number. The lines for marginal instability in the $\rm Rm^*$-$\rm Ha_{\rm crit}$ plane are rather straight lines. The line for $\rm Pm=1$ gives the maximum  stabilization by rigid rotation. In this representation a stronger stabilization does not exist. If ${\rm Rm^*}\lsim 4 \cdot {\rm Ha}$ then all fluids are  unstable even under the influence of   rotation. Note that the rotational stabilization is much weaker for  $\rm Pm\neq 1$. It is particularly  weak for very small $\rm Pm$.  

\begin{figure}
\vskip -3mm
\hskip-10mm
   \includegraphics[width=9.8cm,height=6.7cm]{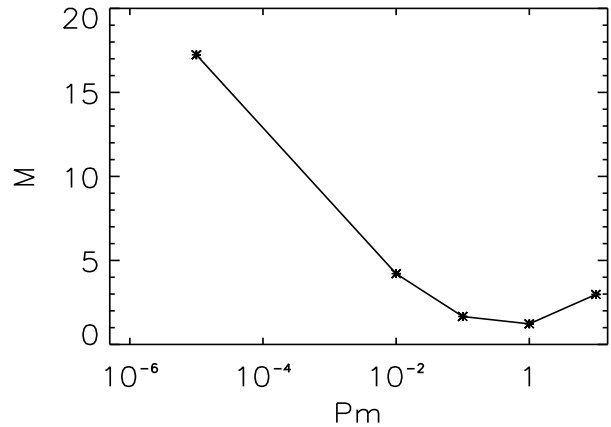}
\vskip-3mm
  \caption{The same as in Fig. \ref{rem} but valid for $\hat\eta\to 0$. Note that fluids with 
 $\rm Pm=1$ can be stabilized with the smallest rotation rates.}  
    \label{M}
 \end{figure}

We find from Fig. \ref{rem} that for fast rotation  the magnetic Mach number (Eq.\,\ref{Mach})
is almost constant for given $\rm Pm$ (with minimum for $\rm Pm=1$) but still depends on the value of $\hat\eta$.  We then also find that the quantity
\beg
\rm M = \sqrt{\hat\eta}\ Mm
\label{Mm}
\ende
does no longer depend on $\hat\eta$ but that it still slightly depends on  $\rm Pm$ (Fig. \ref{M}). From  Eq. (\ref{Mm}) it results that
\beg
\Om = {\rm M}\ \Om_{\rm A}
\label{Mm1}
\ende
is the critical rotation rate which stabilizes  the TI. Here, the Alfv\'{e}n frequency (Eq. \ref{20})
of the {\em outer} magnetic field amplitude has been used.

\begin{figure}
\vskip-4mm
\hskip-2mm
     \includegraphics[width=10.0cm]{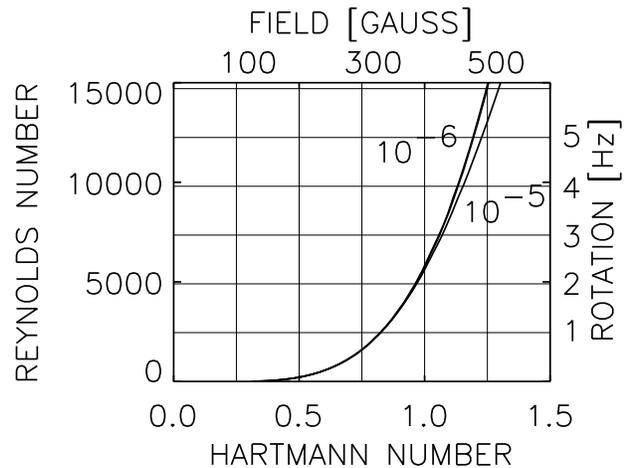}
\vskip-2mm
     \caption{The  suppression of the TI in a wide-gap container by rigid rotation for   $\rm Pm = 10^{-5}$ and $\rm Pm = 10^{-6}$. The standard Reynolds number (\ref{re}) is given for experimental applications. The numerical values of the outer field strength and the critical rotation rates  are given for experiments with Gallistan. $\hat\eta=0.05$,  $\mu_B=20$.}
    \label{f11}
 \end{figure}

The  question is whether     the rotational stabilization of the TI can be probed in the laboratory. The rigidly  rotating wide-gap container with ${\hat\eta=0.05}$ is thus considered also for the very small magnetic Prandtl numbers of fluid metals and for Reynolds numbers of order 10$^3$ where after Fig. \ref{rem} the bifurcation lines are no longer straight lines. The magnetic Prandtl number of sodium is about ${\rm Pm = 10^{-5}}$, and for gallium it is about 
${\rm Pm = 10^{-6}}$. Without rotation the critical Hartmann number is 0.31 for this container (independent of  Pm). With rotation the numerical results  are given in Fig. \ref{f11}. We find the rotational stabilization also existing  for conducting fluids with their very small Pm.  For not too fast    rotation the differences of the resulting critical Ha are very small  for both the fluid conductors   so that  experiments with gallium or sodium are   possible.  For the  small
 Reynolds number of order 1000 the marginal-stable magnetic field is about two times higher than for $\rm Re=0$. With the density (${\rho= 6.4}$\ g/cm$^3$), viscosity ($\nu=3.4\!\times\! 10^{-3}$\  cm$^2$/s) and magnetic diffusivity ($\eta=2428$\ cm$^2$/s) for Gallistan the corresponding magnetic field amplitudes  (at the outer cylinder) in Gauss  and the rotation frequencies (in Hz) are also given in Fig. \ref{f11}. The resulting values are moderate. It should thus not be too  complicated to find the basic effect of  the rotational suppression of TI also   in the MHD laboratory.

Our linear code is  also able to work with ${\rm Pm}=0$. The curves in Fig.  \ref{f11}  do not become more steep for $\rm Pm\to 0$. There is no visible difference between the curves for ${\rm Pm}=10^{-6}$ and ${\rm Pm}=0$.

\begin{figure}
\vskip-5mm
\hskip-10mm
     \includegraphics[width=9.5cm]{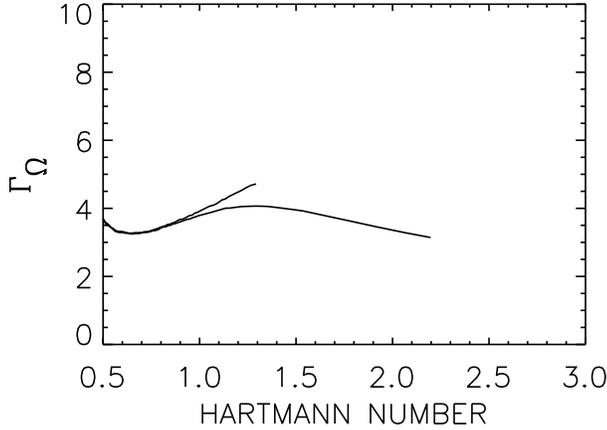}
\vskip-2mm
     \caption{The  parameter $\Gamma_\Omega$  for  $\rm Pm = 10^{-5}$ (long line)  and
 $\rm Pm = 10^{-6}$ (short line).}
    \label{GO}
 \end{figure}

Figure \ref{f11} does not concern to a  gallium column without inner cylinder. However, as demonstrated above the transition from $\hat\eta=0.05$ to $\hat\eta=0$ is easy. We take from Fig. \ref{f11} that close to the critical Hartmann number for resting container
\beg
\rm Re= \gamma_\Omega (Ha- Ha_{\rm crit})^{2.5}.
\label{Re}
\ende
After the usual transformation to $B_{\rm out}$ 
  one finds 
 \beg
\Om \simeq \Gamma_\Omega \ (\frac{B_{\rm out}}{\sqrt{\mu_0\rho}})^{5/2} 
\frac{R_{\rm out}^{1/2}}{{\rm Pm}^{1/4}\eta^{3/2}}
\label{Ome}
\ende
with $\Gamma_\Omega= {\hat\eta}^{11/4}\ \gamma_\omega$. The result (\ref{Ome}) also holds for $\hat\eta\to 0$. Figure \ref{GO}   shows the numerical values of $\Gamma_\Omega$ for the fluids with small magnetic Prandtl numbers.

Using the numerical result $ \Gamma_\Omega\simeq 4$ from Fig. \ref{GO} Eq.
({\ref{Ome}}) yields
 \beg
\Om \simeq 1.36 \, B_{100}^{2.5}  \quad [{\rm s}^{-1}]
\label{Omeg}
\ende
for a container filled with Gallistan and with ${R_{\rm out}=5}$\ cm or, what is the same,  ${f=0.22\, B_{100}^{2.5}}$  as the critical  rotation frequency which marginally stabilizes the TI.
Hence, for (say) 200 Gauss the limit frequency is 1.2 Hz. which is much faster  in comparison to the container with $\hat\eta=0.05$.

\begin{figure}
\vskip-3mm
\hskip-5mm
     \includegraphics[width=8.0cm]{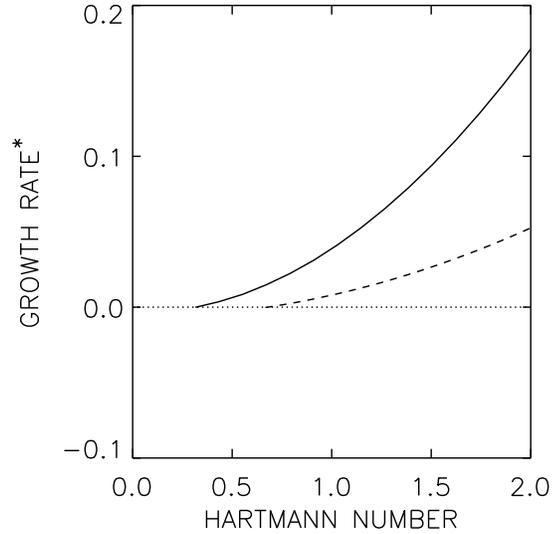}
\vskip-2mm
     \caption{The normalized growth rates for  ${\rm Pm = 10^{-6}}$ under the influence of rigid rotation with ${\rm Re=1000}$ (dashed).   The solid line gives the growth rates for the resting container (see Fig. \ref {pm1}). Note the stabilizing action of global  rotation. $\hat\eta=0.05$,  $\mu_B=20$.}
    \label{f12}
 \end{figure}

The influence of the rotation on the growth rates of the instability is only small.  Figure  \ref{f12} displays for $\rm Pm = 10^{-6}$ the normalized growth rates for ${\rm Re=0}$ (solid line) and $\rm Re=1000$ (dashed line). The stabilizing action of the basic rotation is obvious. 

 \section{The magnetic field  pattern}
  \begin{figure*}[t,b]
    \hbox{
     \includegraphics[width=5.5cm]{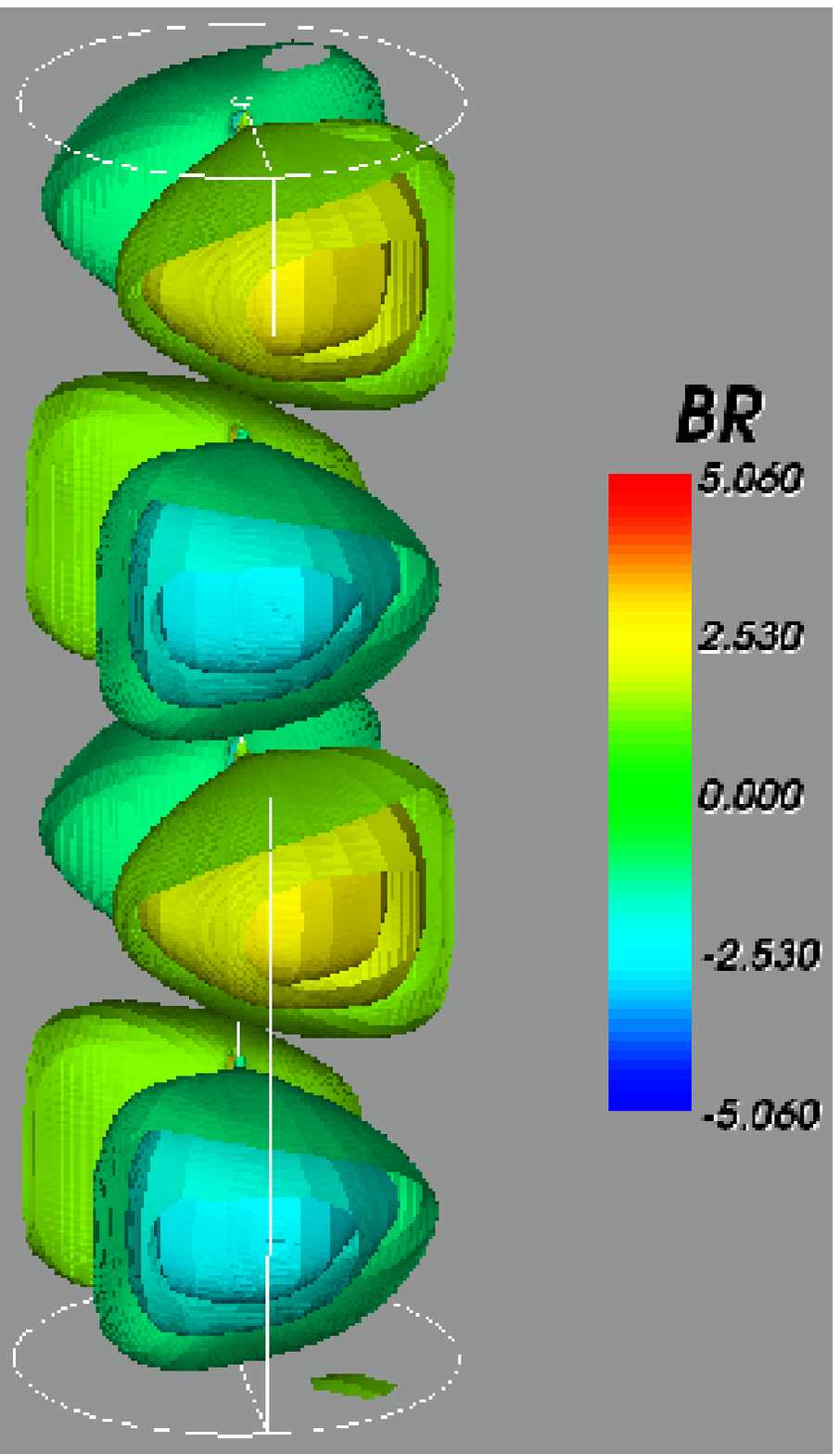}
      \includegraphics[width=5.5cm]{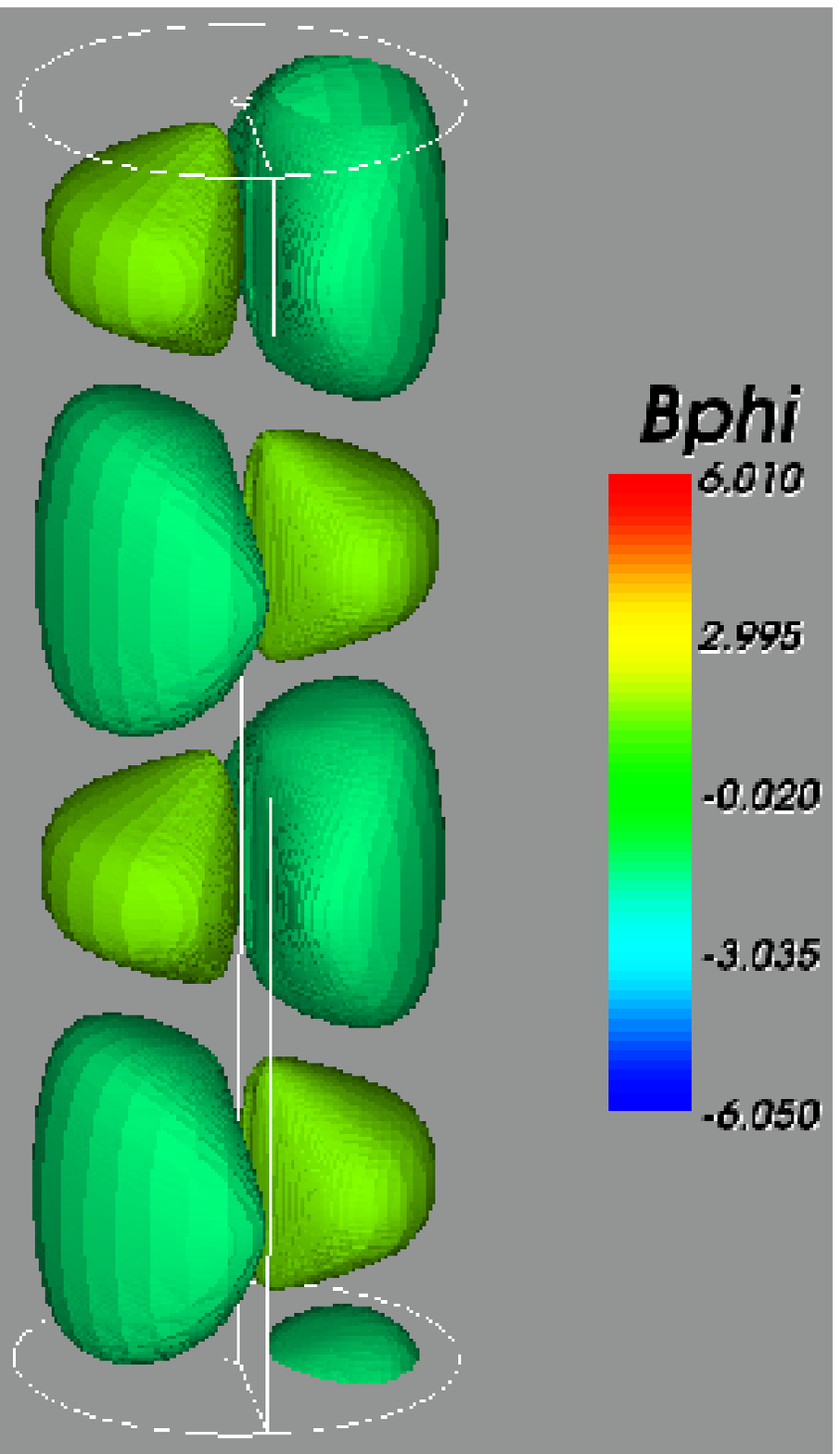}
       \includegraphics[width=5.5cm]{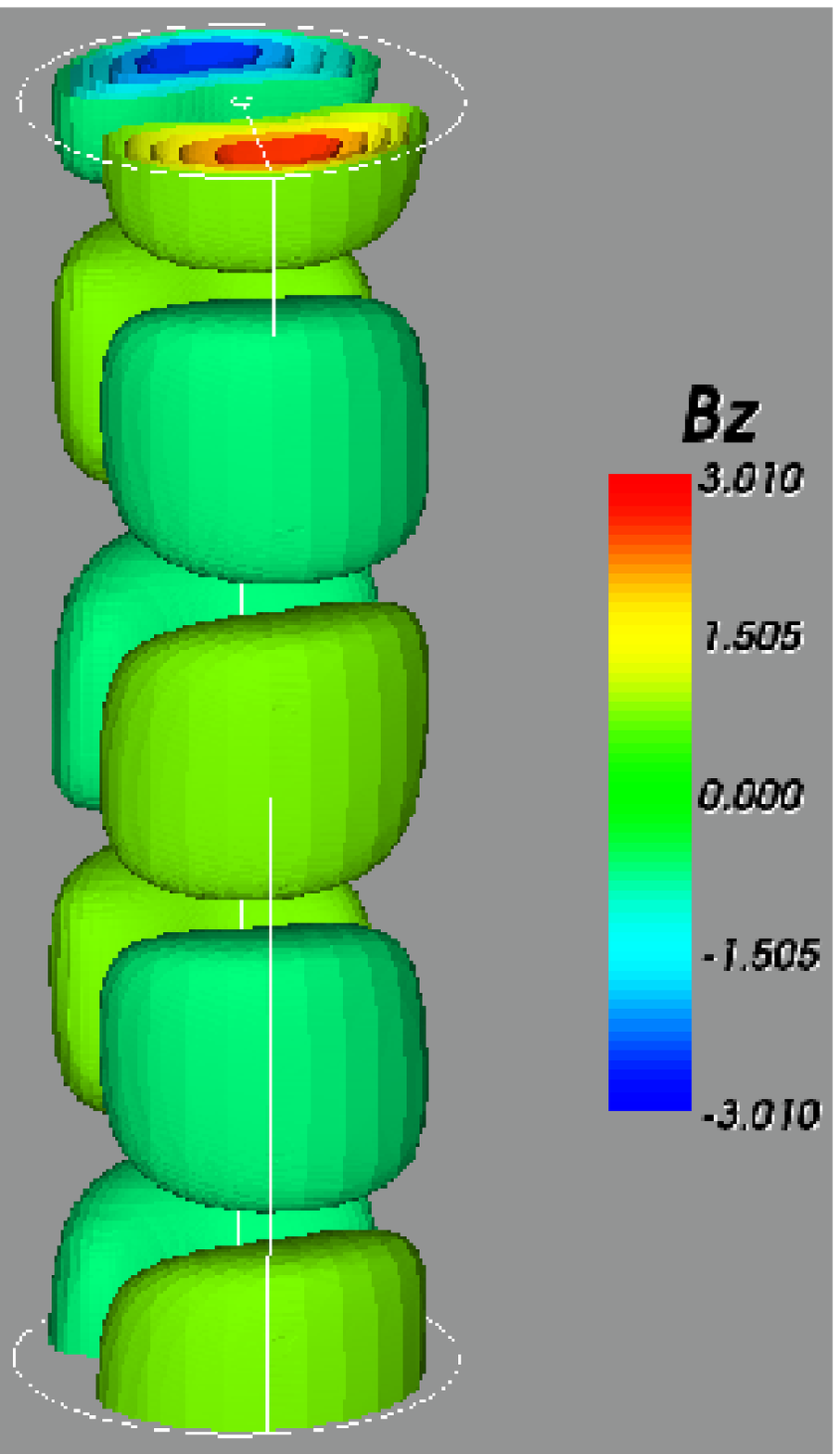}
       }
     \caption{(online colour at: www.an-journal.org) Isosurface plots of all field components. The  outermost visible  surface represents  25\,\% of the maximum value. $\rm Re=0$,  
          $\rm Ha=2$, $\hat\eta=0.05$,  $\mu_B=20$, $\rm Pm=0.02$.}
    \label{f13}
 \end{figure*}
Without rotation the primary current-driven instability leads to a nondrifting steady-state solution. Isosurfaces of
the pattern resulting from a 3D nonlinear simulation are  shown in Fig. \ref{f13} for all the three field components. 
 More details of  the code  which  
solves the incompressible 
MHD equations are given by   Gellert et al. (2008). The boundary conditions at the walls are those   for pseudovacuum.  
The container with  $\hat\eta=0.05$ is  periodic in $z$-direction with a domain height of $2\pi (R_{\rm out}-R_{\rm in})$. The Prandtl number   $\rm Pm=0.02$ is the minimum value  with which the code can operate.  The  Hartmann 
number ($\rm Ha=2$) is supercritical with almost two times the critical value.  The dominating modes with $m=\pm 1$ lead to a  kidney-shaped pattern. Note the different azimuthal 
angles of the pattern of the  azimuthal component compared to the  others.

From  linear calculations we know that for smaller $\rm Pm$  the axial wave numbers decrease and the pattern extensions in axial direction become  larger.

The secondary bifurcation leads to axial  oscillations around  half the wavelength. To
what extent this picture holds for ${\Pm=10^{-6}}$ is open. Because of the $\Pm$-independent critical $\Ha$ and the only slightly
larger Hartmann number used in the simulations shown here, it is possible that liquid metal experiments behave nearly the same.

 \section{A short review}
 
The interplay of toroidal magnetic fields  and solid-body rotation for incompressible fluids of uniform density filling the gap between 
 the cylinders of a Taylor-Couette container is considered. The toroidal field
 is the result of an electric axial
 current of homogeneous density. It is shown that for zero rotation the critical magnetic field amplitudes 
 for marginal Tayler instability 
 do not depend on the magnetic Prandtl number. The critical magnetic field at
 the inner cylinder strongly depends on the gap width.  It is very high for
 small gaps and it this rather low for wide gaps. For small enough inner radius
 $R_{\rm in}$ the critical Hartmann number (of the inner field strength) runs as
 $R_{\rm in}^{1.5}$. Note, however, that  the critical magnetic field at the
 outer cylinder does hardly vary with the gap width (see Eq. \ref{H}). The
 resulting electric currents necessary for TI with $|m|=1 $ slightly exceeds 
 3.2~kA if the material is the same gallium-tin alloy (GaInSn)  as used in the
 experiment PROMISE. Such currents can easily be produced in the laboratory. The
 main difficulty for experiments is given by the  long growth times of the
 instability for magnetic fields weaker than 1 kG.

Also  the rotational quenching of the TI is  studied. Figure \ref{rem} shows the
rotational stabilization   for various magnetic Prandtl numbers. To   normalize 
 the basic rotation a `mixed' Reynolds number (Eq. \ref{rem1}) is used in which  -- as in the Hartmann number -- the viscosities $\nu $ and $\eta$ are symmetric. The ratio of this  Reynolds number  and the Hartmann number (i.e. the magnetic Mach number) is  free of any diffusivity.  This ratio describes the rotational quenching of TI for various Pm. It is shown that the most effective stabilization of TI happens for ${\rm Pm=1}$. It is weaker for both smaller Pm and  higher Pm. Many numerical simulations with $\rm Pm\simeq 1$, therefore, could  overlook the nonaxisymmetric instability of strong toroidal fields.

Our calculations  demonstrate  that rigid rotation always stabilizes the magnetic field against the nonaxisymmetric TI. We have also  shown  that  this rotational stabilization should  be observable in the laboratory. Figure \ref{f11} provides the result that the critical Hartmann number in a wide-gap container of  $\hat\eta=0.05$ can be increased by a factor of two by a rigid rotation of only a few Hz.

\acknowledgements Many discussion with Frank Stefani and Gunter Gerbeth (both
FZD) are cordially  acknowledged about the theory of the current-driven instability and its experimental realization.

\end{document}